\documentclass[preprint,12pt]{elsarticle}
\usepackage[top=1in, bottom=1in, left=1in, right=1in]{geometry}

\usepackage{graphicx}
\usepackage{amssymb}
\usepackage{amsthm}
\usepackage{amsmath}
\usepackage{amsfonts}
\usepackage{multirow}

\usepackage{lineno}
\usepackage{subfigure}

\usepackage[utf8]{inputenc}
\usepackage{tabu}
\usepackage{xcolor}

\makeatletter
\def\ps@pprintTitle{%
   \let\@oddhead\@empty
   \let\@evenhead\@empty
   \let\@oddfoot\@empty
   \let\@evenfoot\@oddfoot
}
\makeatother

\begin{document}

\begin{frontmatter}


\title{Machine Learning the Effective Hamiltonian in High Entropy Alloys \footnote{\footnotesize{This manuscript has been co-authored by UT-Battelle, LLC, under contract DE-AC05-00OR22725 with the US Department of Energy (DOE). The US government retains and the publisher, by accepting the article for publication, acknowledges that the US government retains a nonexclusive, paid-up, irrevocable, worldwide license to publish or reproduce the published form of this manuscript, or allow others to do so, for US government purposes. DOE will provide public access to these results of federally sponsored research in accordance with the DOE Public Access Plan (http://energy.gov/downloads/doe-public-access-plan).}}}



\author{Xianglin Liu$^\dagger$ \corref{cor2}}
\ead{xianglinliu01@gmail.com}
\address{Materials Science and Technology Division, Oak Ridge National Laboratory}

\author{Jiaxin Zhang$^\dagger$ \corref{cor2} 
\footnote{\footnotesize{ $^\dagger$ These two authors contributed equally to this work}}}
\ead{jiaxin.zhanguq@gmail.com}
\address{Center for Computational Sciences, Oak Ridge National Laboratory}

\author{Markus Eisenbach}
\address{Center for Computational Sciences, Oak Ridge National Laboratory}

\author{Yang Wang}
\address{Pittsburgh Supercomputing Center, Carnegie Mellon University}

\begin{abstract}


The development of machine learning sheds new light on the problem of statistical thermodynamics in multicomponent alloys. However, a data-driven approach to construct the effective Hamiltonian requires sufficiently large data sets, which is expensive to calculate with conventional density functional theory (DFT). To solve this problem, we propose to use the atomic local energy as the target variable, and harness the power of the linear-scaling DFT to accelerate the data generating process. Using the large amounts of DFT data sets, various complex models are devised and applied to learn the effective Hamiltonians of a range of refractory high entropy alloys (HEAs). The testing $R^2$ scores of the effective pair interaction model are higher than 0.99, demonstrating that the pair interactions within the 6-th coordination shell provide an excellent description of the atomic local energies for all the four HEAs. This model is further improved by including nonlinear and multi-site interactions. In particular, the deep neural networks (DNNs) built directly in the local configuration space (therefore no hand-crafted features) are employed to model the effective Hamiltonian. The results demonstrate that neural networks are promising for the modeling of effective Hamiltonian due to its excellent representation power.

\end{abstract}


\end{frontmatter}


\section{Introduction}

Density functional theory (DFT) is a powerful method to calculate the properties of materials from first principles. Combined with Monte Carlo methods, DFT not only can be applied to the ground states, but also be used to study finite temperature systems \cite{widom_2018, Eisenbach_2019, Fernandez-Caballero2017}. However, a direct combination of the two is severely hindered by the expensive computational cost. For example, using a supercell of 250-atom \cite{PhysRevB.93.024203}, the ``brute-force" statistical simulation of the order-disorder transition in CuZn alloy calculated 600,000 configurational energies with DFT, which approximately took $10^{8}$ CPU hours. One less ambitious but more practical approach is to establish a surrogate model of the effective Hamiltonian from the DFT data, and use this computationally cheap model in the Monte Carlo simulation. An example of such a strategy is the cluster expansion method \cite{PhysRev.81.988, SANCHEZ1984334, NPJ_Widom, PhysRevB.72.165113, doi:10.1021/ja9105623}, in which the total energy is expressed in terms of the effective cluster interactions (ECIs) \cite{RubanECIReview}. The EPI parameters are typically calculated with the Connolly and Williams approach (also known as the structure inversion method) \cite{PhysRevB.27.5169} from the DFT energies of a set of selected structures. While cluster expansion provides a systematic approach to the construction of the effective Hamiltonian, its application to complex systems remains challenging due to rapid increase of the number of ECIs with respect to the chemical components \cite{PhysRevLett.116.105501, seko2009cluster, widom_2018}. This is particularly true for a class of novel materials known as high entropy alloys (HEAs) \cite{ADEM:ADEM200300567, CANTOR2004213, George2019}, which contain more than four principal elements, and demonstrate some exceptional mechanical properties \cite{NatureComNiCoCr, Gludovatz1153, SENKOV2011698, MIRACLE2017448}. Other than cluster expansion, linear response methods based on the coherent potential approximation (CPA), such as the $S^{(2)}$ theory \cite{PhysRevLett.50.374}, the generalized perturbation method (GPM) \cite{0305-4608-6-11-005}, and the embedded-cluster method (ECM) \cite{PhysRevB.36.4630}, are also widely used to calculate the ECIs. The advantage of these methods is that they directly relate the ECIs to the underlying electronic structure, while the shortcoming is that they are limited by the single-site approximation.

A different strategy to construct the effective Hamiltonian is through supervised machine learning \cite{ML_Cluster, 2015arXiv151209110G}. Compared to the traditional structure inversion method, machine learning adopts a data-driven approach that generally requires large data sets. Moreover, machine learning models are generally more complex, with large numbers of parameters, therefore in principle possess better representation capability than simple models. In supervised machine learning, the training of model is achieved through adjusting the parameters so as to minimize the loss function of the training data. Machine learning has demonstrated great success in a wide range of fields, such as computer vision and natural language processing \cite{NatureDL}. In materials science, machine learning has been applied to model physical quantities such as atomic forces \cite{Chmielae1603015, PhysRevLett.114.096405}, interatomic potentials \cite{PhysRevB.95.094203, PhysRevX.8.041048, 2015arXiv151209110G, Korman_npj}, and formation energies \cite{ML_Cluster, Ye2018}.

While increased model complexity generally improves the representation capability, it also makes the model sensitive to random noise. This phenomenon is known as the bias-variance trade-off \cite{MEHTA20191}. Generally speaking, sufficiently large training data sets are required for the complex model to demonstrate better performance (smaller out-of-sample error) than the simple ones. As a result, for machine learning the Hamiltonian, it is critical to generate data efficiently so that more accurate models can be employed. In fact, the availability of large data sets, and the capability to handle them, is one of the most important reasons for the resurgence of deep neural networks (DNN) in the last 10 years \cite{Goodfellow-et-al-2016}. However, as previously mentioned, DFT methods are computationally expensive, and this is the exact reason that we resort to the surrogate model instead of directly using DFT. At first sight, the analysis seems to indicate a dilemma that huge amounts of DFT calculations are inevitable even for the surrogate model approach.

\begin{figure}[!ht]    
    \centering
\includegraphics[width=0.4\textwidth]{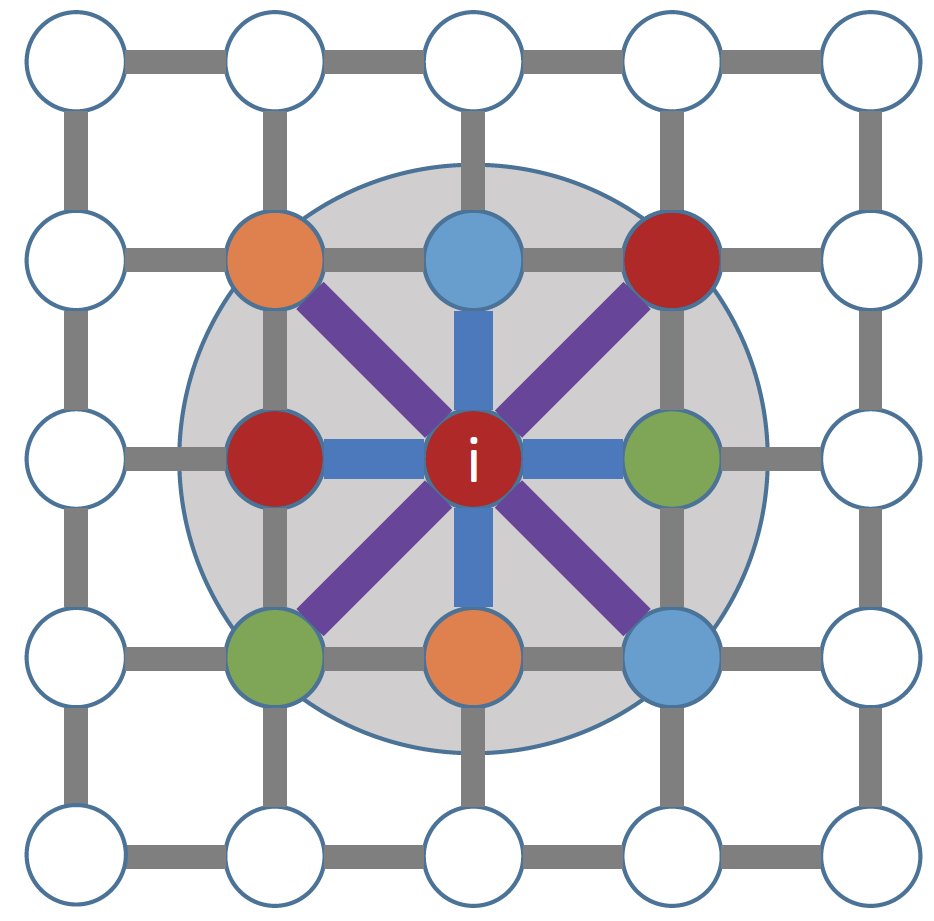}
  \caption{ (Color online) A schematic to illustrate the dependence of the local energy $E_i$ on the local chemical environment. The grey circle demonstrates the local interaction region, with the local atom in the center. In practice, the local interaction zone for each atom includes multiple neighboring shells. The different colors of the atoms signify different chemical species. Pair interactions of the nearest neighbor and next nearest neighbor are also shown.} 
\label{fig:EPI_schematic}
\end{figure}

In this work, we propose to solve the above problem by a combination of two ideas: linear-scaling DFT \cite{RevModPhys.71.1085} and atomic local energies. Linear-scaling DFT utilizes the nearsightedness principle \cite{Prodan11635} to reduce the computational complexity of conventional DFT from $O(N^3)$ to $O(N)$, where $N$ is the number of atoms in the supercell. Compared to conventional DFT, the speed advantage of linear-scaling is particularly significant for large supercells. Different linear-scaling DFT methods have been implemented. In this work, we focus on the locally self-consistent multiple scattering (LSMS) method \cite{PhysRevLett.75.2867}. The LSMS method achieves linear scaling by restricting the quantum scattering of electrons within the so called local interaction zone, but it still evaluates the electrostatic interactions everywhere. In the LSMS method, the concept of atomic local energy is actually very natural since it always divides the space into Voronoi polyhedra corresponding to each atom. The atomic local energy can thus be obtained by integrating the energy density within the polyhedron, at a negligible computational cost. By using the local chemical environment as the input and atomic local energy as the output, as illustrated in Fig.~\ref{fig:EPI_schematic}, a single DFT calculation with an $N$-atom supercell generates $N$ data sets. Therefore, the combination of linear-scaling DFT and atomic local energies technically reduces the time scaling of generating one DFT data from $O(N^3)$ to $O(0)$, which renders the use of complex models with thousands of parameters feasible. 

\section{Results and discussion}
In general, the effective Hamiltonian can be obtained from two different approaches. One is to first project the local chemical environment to the space spanned by a series of physical features, such as ECIs, then establish a mapping from the features to the atomic local energy. Note that unlike the traditional cluster expansion, the mapping between ECIs and energy is not necessarily linear. A more interesting approach is to seek a mapping directly from the local environment to the atomic local energy. Without the intermediate dimension reduction, this method is more accurate in principle, and can be easily generalized to other materials. Of course, in both cases a cutoff has to be set on the interaction range, otherwise the feature space will be of infinite dimension. In the first approach, the local energy of the $i$-th lattice site can be written as
\begin{align}
E_i = H (\vec{\Sigma}_i ) + \epsilon,
\end{align}
where  $\epsilon$ denotes the random error due to the approximations in the model, and $\vec{\Sigma} = (\Sigma^1, \Sigma^2, \cdots, \Sigma^{N_f})$ are the $N_f$ physical features calculated from the local chemical environment. In the second approach, the local energy is given by
\begin{align}
E_i = H (\vec{\sigma}_i ) + \epsilon,
\end{align}
where $\vec{\sigma} = (\sigma^0, \sigma^1, \cdots, \sigma^{N_{L}-1})$ is a configuration of $N_L$ atoms in the local interaction region (LIR), with $\sigma^0$ representing the local atom. After obtaining the local energies, the total energy is simply given by summing them over all lattice sites, i.e.,
\begin{align}
E_t = \sum_i E_i.
\end{align}

\subsection{Linear EPI}
\begin{figure}[h!]
\centering
   \includegraphics[width=1.0\textwidth]{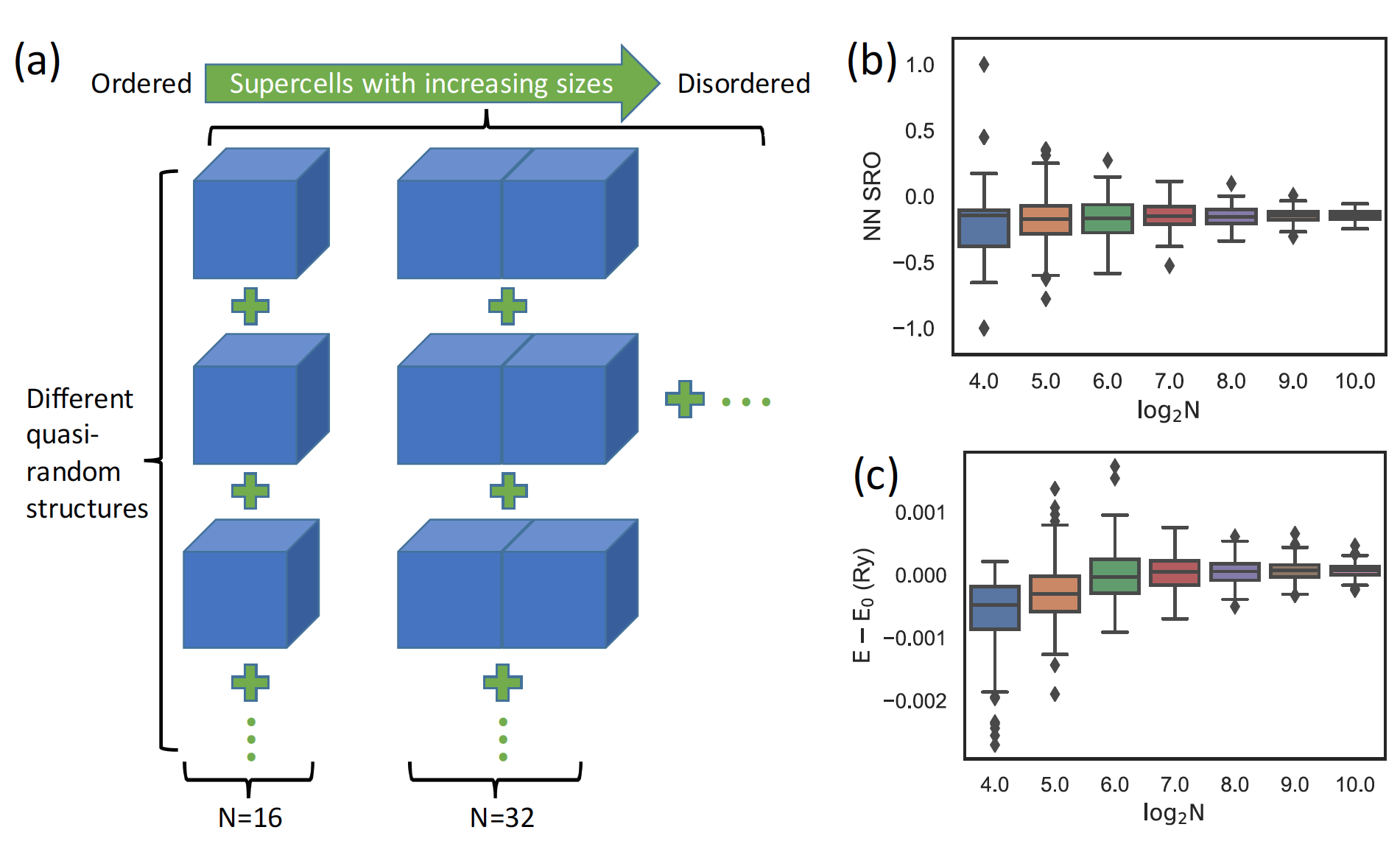}
\caption{(Color online) (a) A schematic of to illustrate our method to obtain the training data set. (b) Box plot of the nearest-neighbor short range order (SRO) parameters in the MoNbTaW data sets, for different supercell size N. (b) Box plot of the averaged total energies in the MoNbTaW data sets, for different supercell size N. } \label{fig:train}
\end{figure}

Different models of the effective Hamiltonian are investigated in this work. We first start with the linear effective pair interaction (EPI) model, in which the local energy is given by a summation of all the pair interactions within the local interaction region,
\begin{align}
E_i = \sum_{f} V^f \Pi^f(\vec{\sigma}_i)  + V^0 + \epsilon,
\label{Linear_EPI}
\end{align}
where $V^f$ are the EPI parameters and $\Pi^f$ are the number of pair interactions of type $f$. The feature index $f$ is actually made up of three parts $(p, p', m)$, representing the element of the local atom, the element of the neighboring atoms, and the coordination shell, respectively. It is obvious that the EPI parameters in Eq.~\ref{Linear_EPI} can be obtained with linear regression. The number of data sets is $N \times N_c$, where $N_c$ is the number of configurations. In practice, the data sets of an $n$-component system are divided into $n$ parts according to the elements,  with one model fitted for each chemical component. To improve the representativeness of the data, the training sets need to contain various order and disorder structures. One simple approach is to carry out the DFT calculation using supercells of different sizes, as illustrated in Fig.~\ref{fig:train}. Within each supercell, the atoms are still randomly distributed, but due to the periodic boundary conditions, configurations from small supercells naturally contains a range of ordered structures, while configurations from large supercells are close to random states. 

\begin{figure}[!ht]    
    \centering
\subfigure[]{\includegraphics[width=0.4\textwidth]{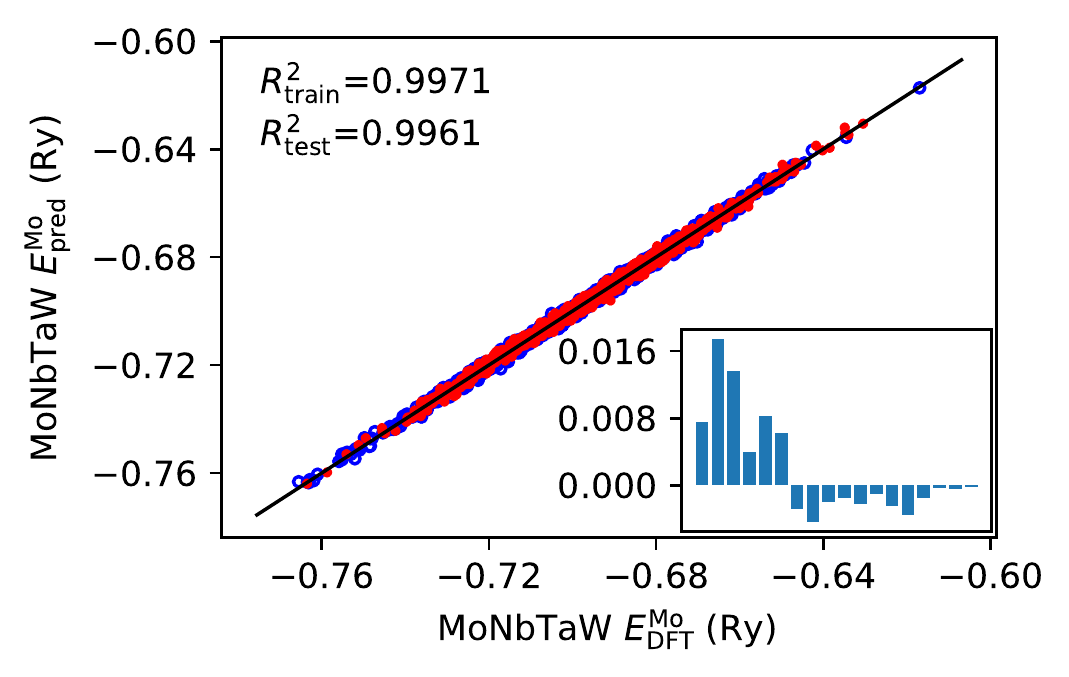}}
\subfigure[]{\includegraphics[width=0.4\textwidth]{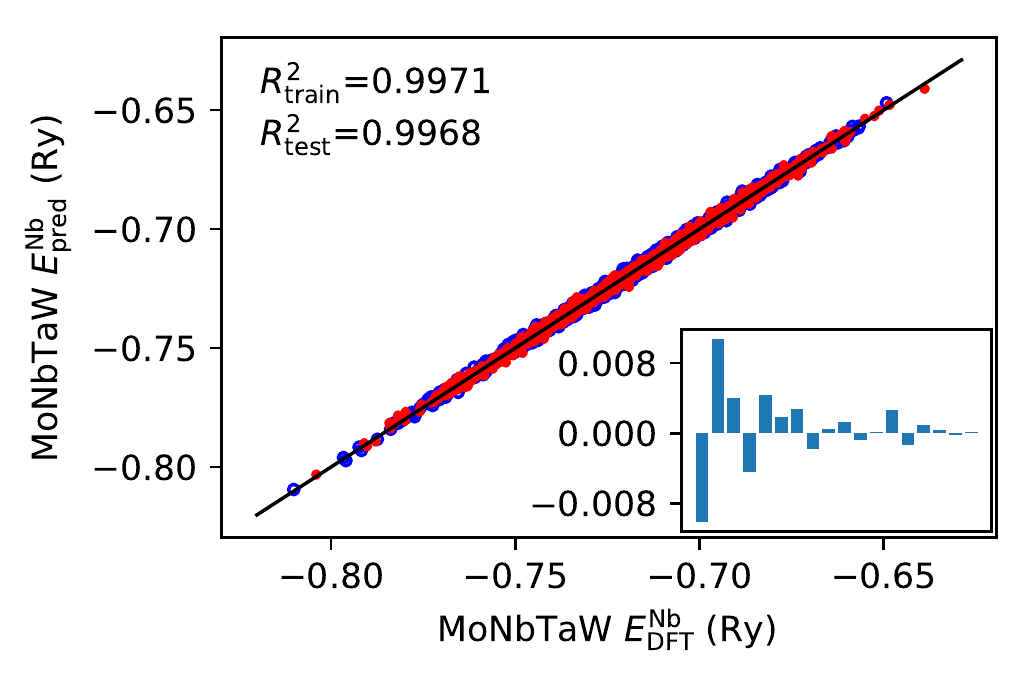}}
\subfigure[]{\includegraphics[width=0.4\textwidth]{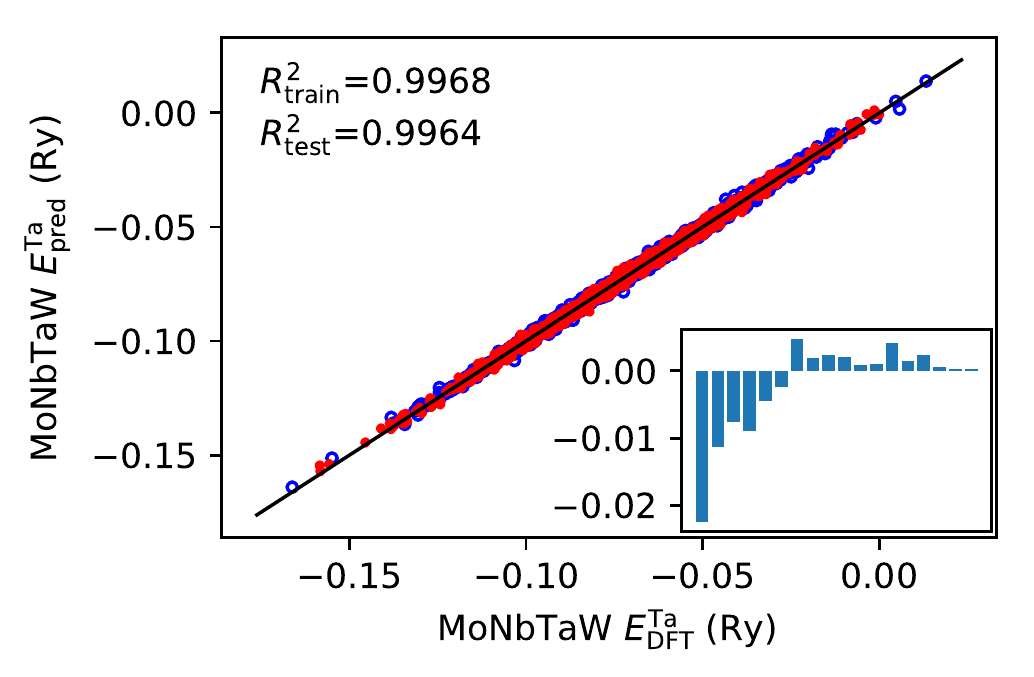}}
\subfigure[]{\includegraphics[width=0.4\textwidth]{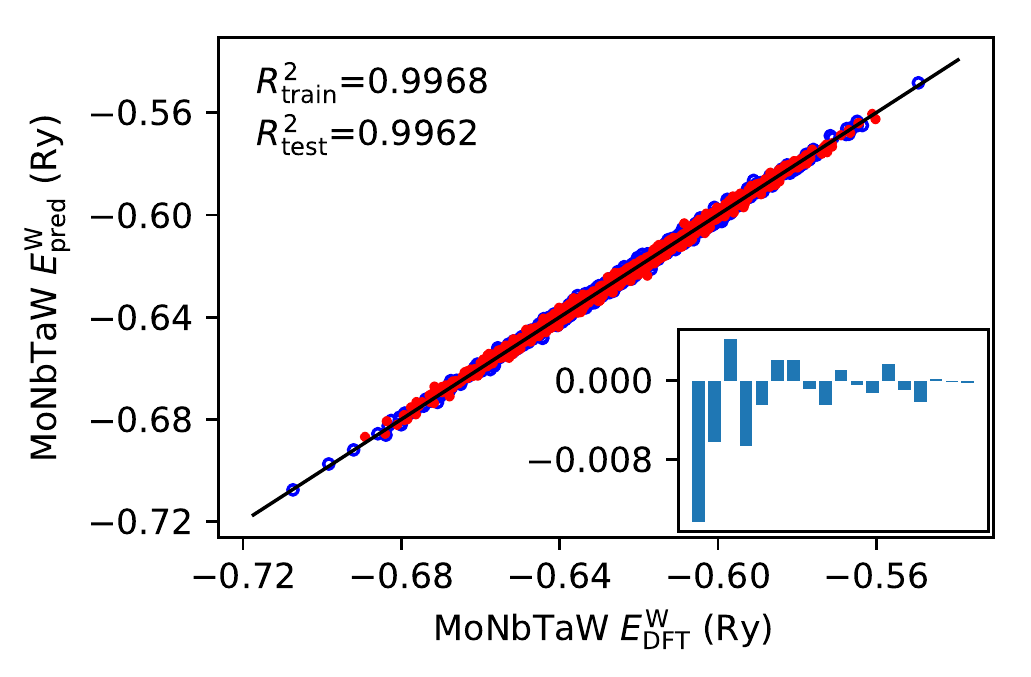}}
    \caption{(Color online) Comparison of the local energy predicted by the linear EPI model with the ones from DFT. The blue circles represent the training data and the filled red circles represent the testing data. The EPI parameters are shown in the bar plots. (a) Mo, (b) Nb, (c) Ta, and (d) W.}  \label{fig:EPI_MoNbTaW}
\end{figure}

The EPI model with the number of coordination shells $m_{\rm{max}} = 6$ is applied to the MoNbTaW refractory HEA and the linear regression results are shown in Fig~\ref{fig:EPI_MoNbTaW}. It is easy to see that this model gives a very accurate prediction of the atomic local energies, with testing $R^2$ scores higher than 0.996 for all the elements. The fitted EPI parameters are also shown in Fig~\ref{fig:EPI_MoNbTaW}, and are arranged according to the $p’$ and $m$ indices. Note that due to the constraint that the total number of atoms are fixed for each coordination shell, only $n - 1$ $p’$ indices are independent for an $n$-element system, therefore we adopt the notation that $p’ \neq p$ and there are a total of $(n-1)\times m_{\rm{max}} $ parameters for each element. For instance, the 18 EPI parameters in Fig~\ref{fig:EPI_MoNbTaW}(a) are arranged according to the feature indices (Mo, Nb, 1), (Mo, Ta, 1), (Mo, W, 1), $\cdots$, (Mo, W, 6). It is also easy to see that the nearest neighbor MoTa pair are the strongest pair interaction, which is in agreement with the results in literature \cite{Huhn2013, ZHANG2020108247}. The coordination shell cutoff $m_{\rm{max}}$ determines the size of the local interaction region. The impact of different $m_{\rm{max}}$ on the root mean square errors (RMSEs) are shown in Fig.~\ref{fig:EvsShell}. It can be seen that the pair interactions within the first 5 shells contribute significantly to the local energies of MoNbTaW, and the effects from longer pair interactions are small. Other than MoNbTaW, we also apply the EPI model to the other three refractory HEAs and the results are listed in Tab.~\ref{tab:t2}. For all the four materials, the testing $R^2$ scores of the local energies are higher than 0.99, demonstrating that the EPI model is generally a good description of the configuration energies. The $R^2$ score of MoNbTaTiW and AlMoNbTaW are a little lower than MoNbTaW and MoNbTaVW, indicating that the addition of Ti and Al introduces stronger high-order interactions into the system.

\begin{figure}[h!]
\centering
   \includegraphics[width=0.6\textwidth]{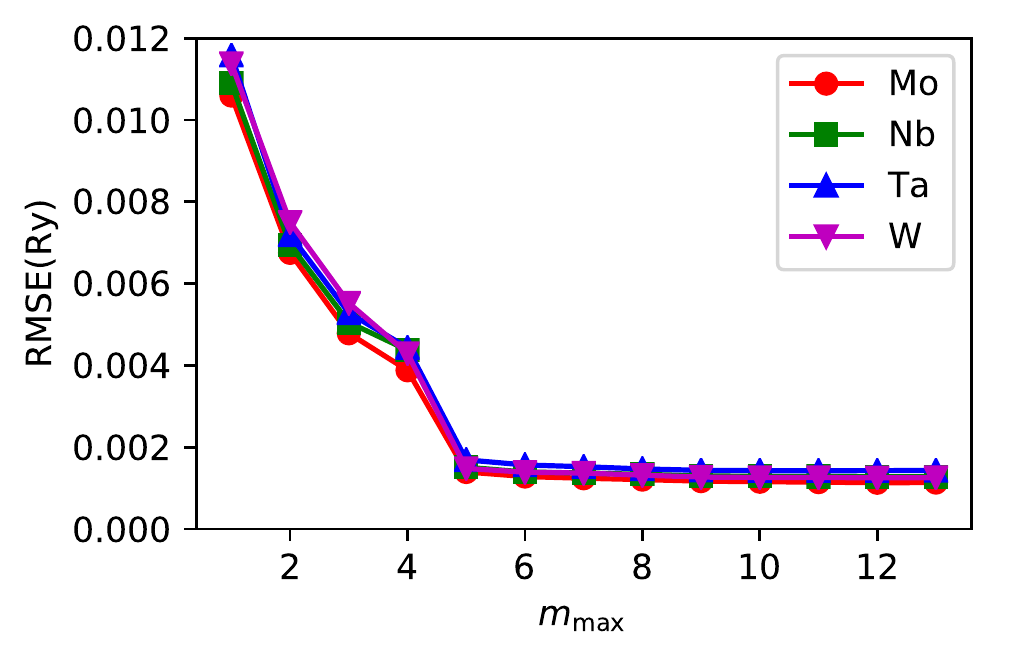}
\caption{(Color online) The dependence of the testing RMSE on the maximum number of coordination shells, for the linear EPI model of MoNbTaW. } \label{fig:EvsDatal}
\end{figure}

\begin{table}[!ht] 
\centering
\caption{Summary of the testing $R^2$ scores, root mean square errors (RMSE), and maximum absolute deviations (MAD) for four different refractory HEAs. Both linear regression (LR) and quadratic regression (QR) of the EPI model are employed. }
\label{tab:t2}
\begin{tabular}{cc|cc|cc|cc }
\hline
\hline
Material & Element & LR-$R^2$ & QR-$R^2$ & LR-RMSE & QR-RMSE & LR-MAD  & QR-MAD \\ \hline
\multirow{4}{*}{MoNbTaW} & Mo&0.99624 &0.99668 &0.00131 &0.00123 &0.00684 &0.00566   \\
 & Nb &0.99647 &0.99675 &0.00142 &0.00136 &0.00592 &0.00539 \\
 & Ta &0.99640 &0.99675 &0.00153 &0.00145 &0.00707 &0.00642\\
 & W &0.99632 &0.99673 &0.00140 &0.00132 &0.00683 &0.00609\\ \hline
\multirow{5}{*}{MoNbTaVW} 
& Mo &0.99852 &0.99882 &0.00208 &0.00186 &0.00935 &0.00762 \\
& Nb &0.99823 &0.99874 &0.00233 &0.00197 &0.01401 &0.00839 \\
&Ta &0.99818 &0.99876 &0.00249 &0.00206 &0.01363 &0.00911 \\
& V &0.99819 &0.99882 &0.00212 &0.00171 &0.01445 &0.01251 \\
& W &0.99852 &0.99885 &0.00225 &0.00198 &0.01072 &0.00874 \\ \hline
\multirow{5}{*}{MoNbTaTiW} 
& Mo &0.99480 &0.99575 &0.00294 &0.00266 &0.01848 &0.01312 \\
& Nb &0.99203 &0.99389 &0.00356 &0.00311 &0.02696 &0.02017 \\
& Ta  &0.99141 &0.99331 &0.00385 &0.00340 &0.02273 &0.01988 \\
& Ti &0.98915 &0.99113 &0.00382 &0.00345 &0.02310 &0.02293 \\
& W &0.99475 &0.99561 &0.00315 &0.00288 &0.01975 &0.01372 \\ \hline
\multirow{5}{*}{AlMoNbTaW} 
& Al &0.99535 &0.99645 &0.00393 &0.00343 &0.02231 &0.01729 \\
& Mo &0.99075 &0.99299 &0.00583 &0.00508 &0.02472 &0.02684 \\
& Nb &0.99043 &0.99187 &0.00573 &0.00529 &0.02724 &0.02584 \\
& Ta &0.99054 &0.99203 &0.00592 &0.00543 &0.02908 &0.03731 \\
& W &0.99107 &0.99317 &0.00604 &0.00528 &0.03114 &0.03290\\ \hline
\end{tabular}
\end{table}

\subsection{Quadratic EPI}
\begin{figure}[h!]
\centering
   \includegraphics[width=0.6\textwidth]{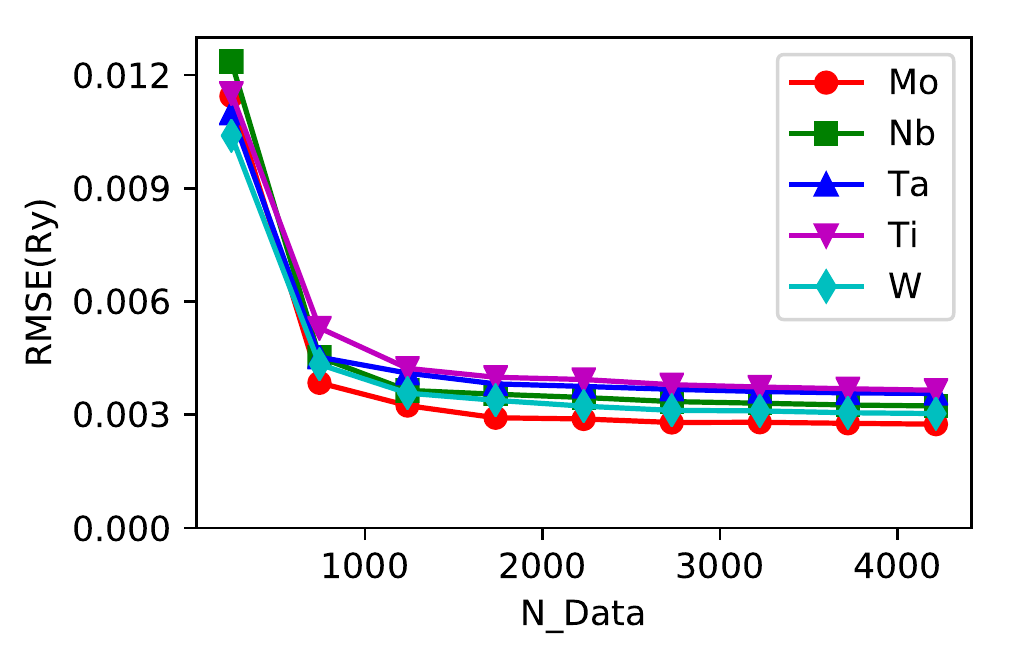}
\caption{(Color online) The dependence of the testing RMSE on the number of data sets for each element. The results are from the quadratic EPI model of MoNbTaW. The data points correspond to 248, 744, 1240, 1736, 2232, 2728, 3224, 3720, and 4216 data sets, respectively.} \label{fig:EvsShell}
\end{figure}

The linear EPI model can be improved by introducing interaction terms with quadratic regression, where the local energy can be written as
\begin{align}
E_i = \sum_{f} V^f \Pi^f(\vec{\sigma}_i)  + \sum_{f,f'\leq f} V^{ff'} \Pi^f(\vec{\sigma}_i) \Pi^{f'}(\vec{\sigma}_i) + V^0 + \epsilon.
\label{QuadraticEPI}
\end{align}
The addition of quadratic terms increases the dimension of the feature space from $k = (n-1)\times m_{\rm{max}} $ to $k \times(k+3)/2$. Using the MoNbTaTiW HEA as an example, the 6-shell quadratic EPI model thus has a total of 324 feature parameters. To avoid overfitting, these parameters are determined by ridge regression, with the L2 regularization parameter taken as $\alpha = 1.0$. The impact of the data set size on the RMSE of MoNbTaTiW is shown in Fig.~\ref{fig:EvsShell}, from which we can see that thousands of data sets are needed for the quadratic EPI model to converge. The results for all the four refractory HEAs are also shown in Tab.~\ref{tab:t2}. It is easy to see that the quadratic models indeed demonstrate better performance than the linear models.

\subsection{Configuration space}
In contrast to the EPI models that make use of the lattice symmetry for dimension reduction, the following models work directly on the chemical configuration space. In the simplest case, one can express the atomic local energies as a linear summation of contributions from each lattice sites in the local interaction region, i.e.,
\begin{align}
E_i = \sum_{j\in \rm{LIR}} V^j \sigma_i^j  + V^0 + \epsilon.
\label{Linear_Config}
\end{align}
Note that $\sigma_i^j$ are categorical variables and an embedding procedure is needed to represent them in real space. Since each lattice site has to be occupied by one of the $n$ elements, we choose to use $n-1$ dummy variables to encode $\sigma_i^j$ . For example, the representations of Mo, Nb, Ta, Ti, W in the MoNbTaTiW HEA are (0, 0, 0, 0), (1, 0, 0, 0), (0, 1, 0, 0), (0, 0, 1, 0), and (0, 0, 0, 1), respectively. As a result, for $m_{\rm{max}}=6$ (64 atoms within the LIR), the number of features for one element increases from 24 in the linear EPI model to 256. Due to lattice symmetry, $V^j$ of the same bonding type should have the same value, so this configuration space linear model is essentially the same as the linear EPI model. However, the benefits of the configuration space representation are that the complete information about the local chemical environment is preserved, and the inclusion of higher order interactions is straightforward. For example, adding quadratic interactions to the nearest neighbors, the local energies can be written as
\begin{align}
E_i = \sum_{j\in \rm{LIR}} V^j \sigma_i^j +\sum_{j,k>j\in \rm{LIR}} V^{jk} \sigma_i^j \sigma_i^k + V^0 + \epsilon,
\label{Quadratic_Config}
\end{align}
which amounts to include all the triplet interactions within the first coordination shell. Using MoNbTaTiW as an example, the calculation results of this model (1ST) are shown in Tab.~\ref{tab:t3}. Compared to the data in Tab.~\ref{tab:t2}, it is easy to see that this model performs better than the linear EPI model, but the improvement is very small, indicating that these triplet interactions are not important for the MoNbTaTiW HEA.

\begin{table}[!ht] 
\centering
\caption{Summary of the testing $R^2$ score, root mean square error (RMSE), and maximum absolute deviation (MAD) for the configuration space models of MoNbTaTiW. 1ST stands for the first-shell-triplet model and NN stands for the neural network model. }
\label{tab:t3}
\begin{tabular}{cc|cc|cc|cc }
\hline
\hline
Material & Element & 1ST-$R^2$ & NN-$R^2$ & 1ST-RMSE & NN-RMSE & 1ST-MAD  & NN-MAD \\ \hline
\multirow{5}{*}{MoNbTaTiW} 

&Mo&0.99505 &0.99306 &0.00287 &0.00340 &0.01341 &0.01706 \\
&Nb&0.99292 &0.98823 &0.00335 &0.00432 &0.02200 &0.02547 \\
&Ta&0.99263 &0.98977 &0.00357 &0.00420 &0.01863 &0.02237 \\
&Ti&0.98980 &0.98703 &0.00370 &0.00417 &0.02432 &0.03418 \\
&W&0.99485 &0.99305 &0.00312 &0.00363 &0.01454 &0.02421
\\ \hline
\end{tabular}
\end{table}

\subsection{Neural Network Model}
\begin{figure}[h!]
\centering
   \includegraphics[width=0.6\textwidth]{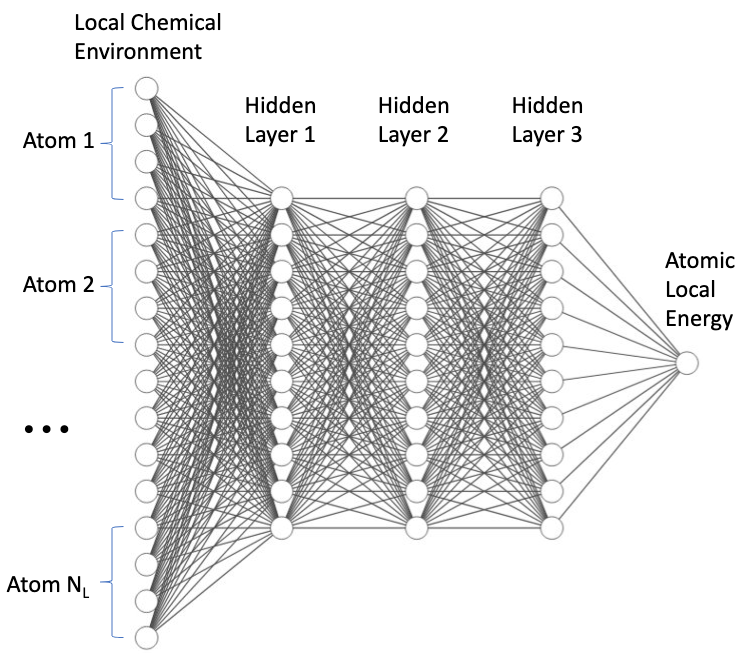}
\caption{ A schematic of the neural networks used to model the effective Hamiltonian. The input layer is made up of dummy variables representing the local chemical environment. } \label{fig:NN}
\end{figure}
\begin{figure}[!ht]    
    \centering
\includegraphics[width=0.8\textwidth]{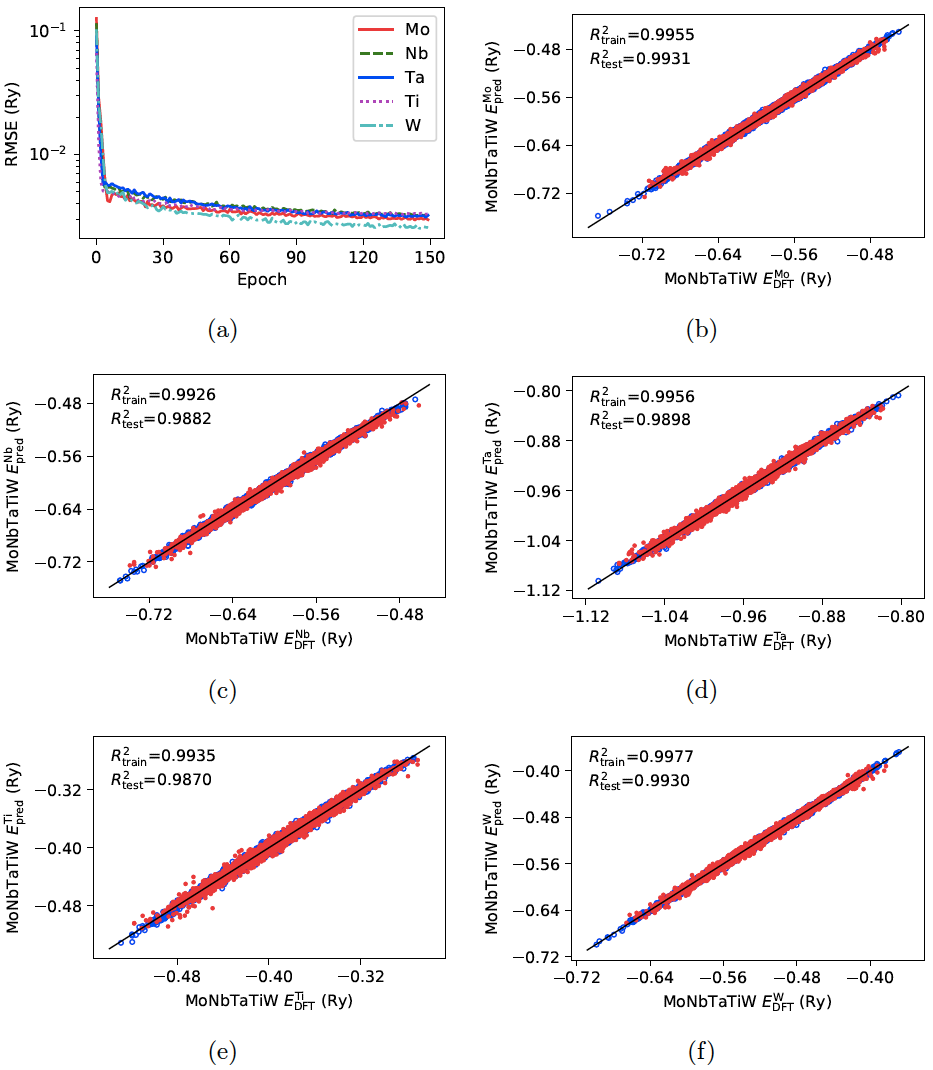}
    \caption{(Color online) (a) Convergence of MoNbTaTiW RMSE with respect to the number of epochs. (b)-(f) Comparison of the local energy predicted by the neural network model with the ones from DFT. The blue circles represent the training data and the filled red circles represent the testing data. (b) Mo, (c) Nb, (d) Ta, (e) Ti, and (f) W. }  \label{fig:NN_Energy}
\end{figure}
\begin{figure}[!ht]    
    \centering
\subfigure[]{\includegraphics[width=0.4\textwidth]{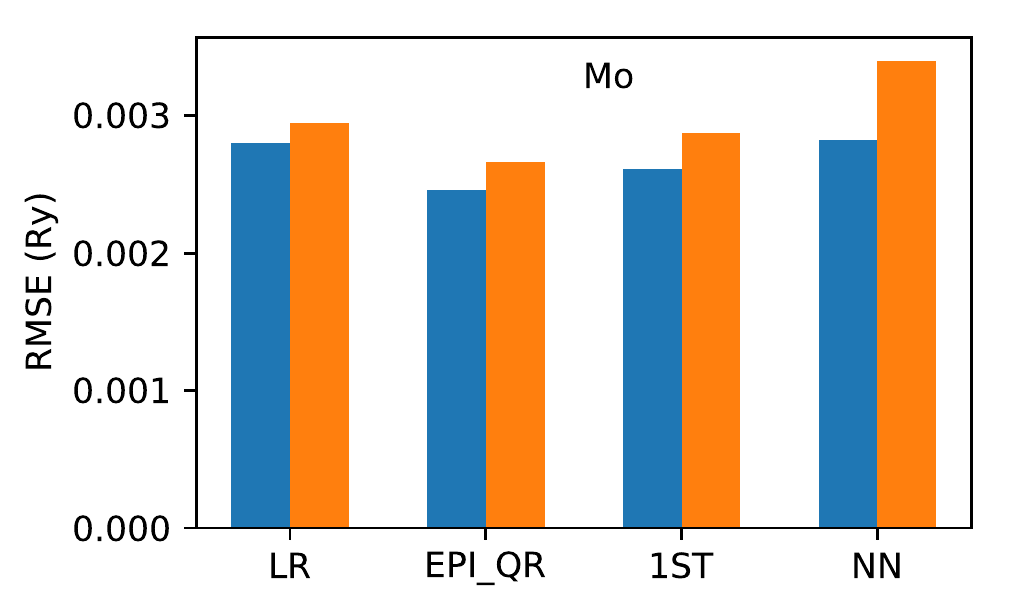}}
\subfigure[]{\includegraphics[width=0.4\textwidth]{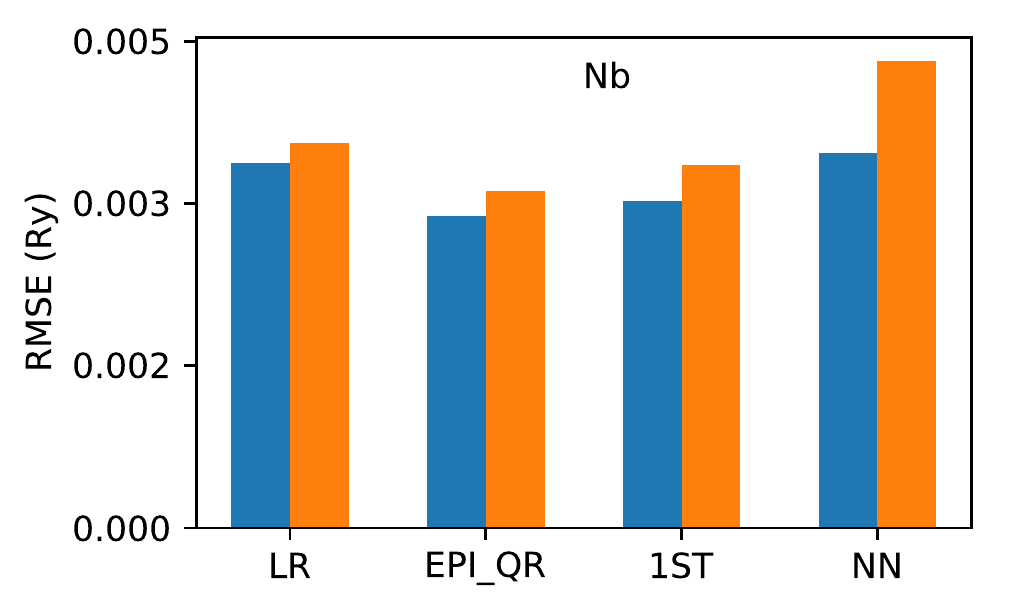}}
\subfigure[]{\includegraphics[width=0.4\textwidth]{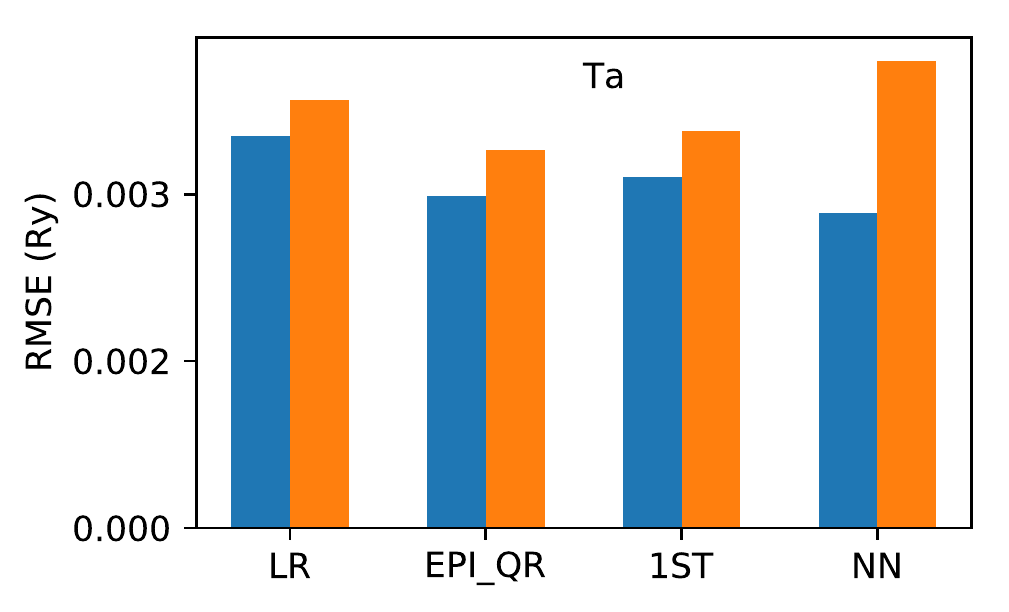}}
\subfigure[]{\includegraphics[width=0.4\textwidth]{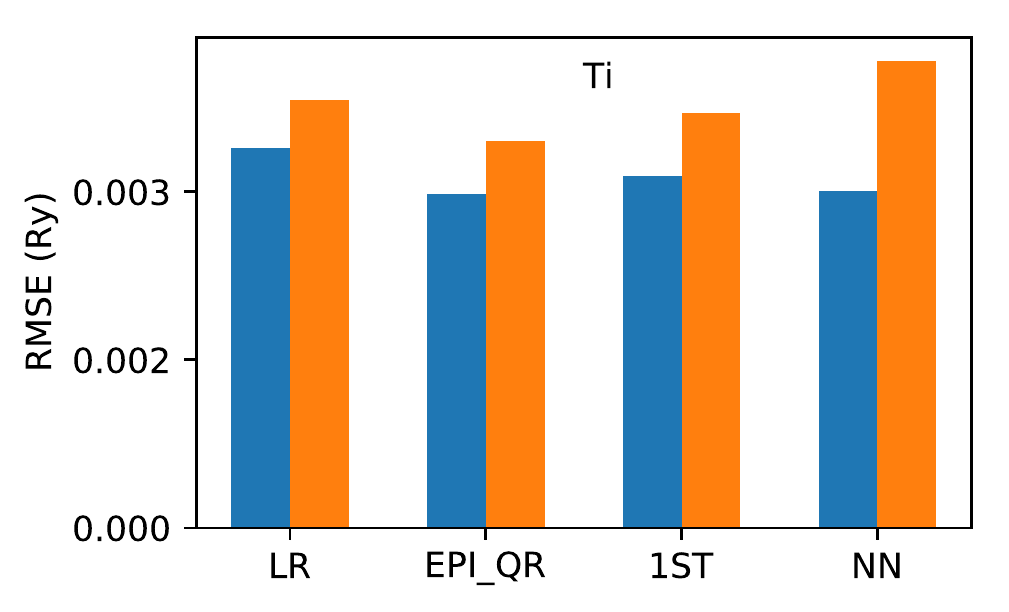}}
\subfigure[]{\includegraphics[width=0.4\textwidth]{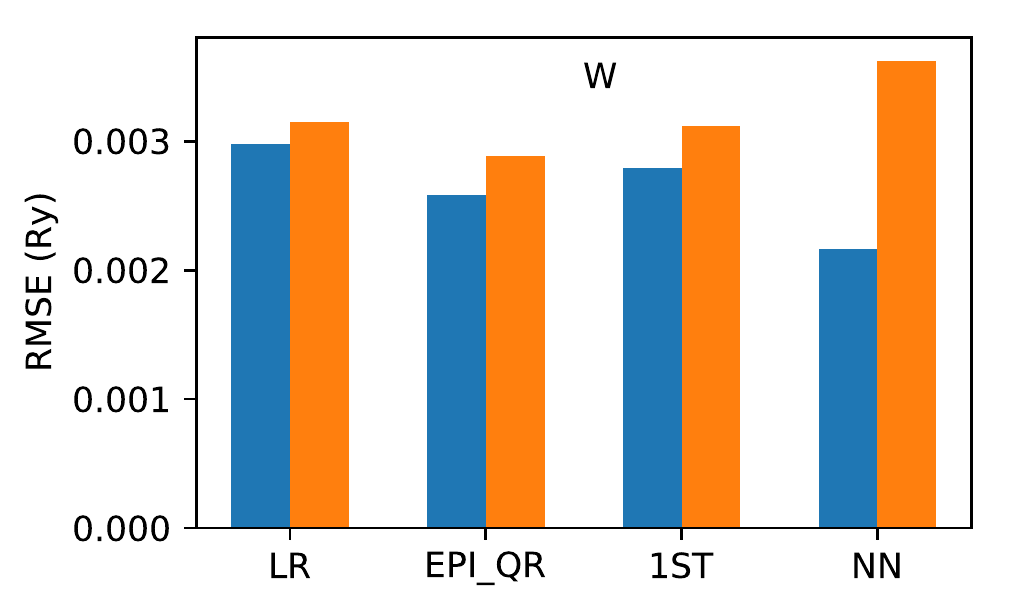}}
    \caption{(Color online) Comparison of the training (blue) and testing (orange) RMSEs from different models. LR stands for the linear EPI model. EPI{\_}QR stands for the quadratic EPI model. 1ST represents the first shell triplet model and  NN denotes the neural network model. }  \label{fig:Model_Comparison}
\end{figure}

Deep neural networks (DNNs) provide a general solution to the problem of constructing effective Hamiltonian in the configuration space. This is a very attractive approach because it involves no ``handcrafted" features and is exact in principle, according to the universal approximation theorem \cite{LESHNO1993861}. A schematic of the neural network architecture employed in this work is shown in Fig.~\ref{fig:NN}, and applied to the MoNbTaTiW HEA. In practice, 20 neurons are adopted for each of the 3 hidden layers, where the rectified linear (ReLU) activation functions are used. The input layer is made up of 256 dummy variables ($m_{\rm{max}} = 6$) and the calculation details are described in the method section. The total number of epochs is 150, and the convergence of RMSE with respect to the epoch number are shown in Fig.~\ref{fig:NN_Energy}(a). The $R^2$ scores, RMSEs, and MADs are listed in Tab.~\ref{tab:t3} and the comparisons between the predicted energy and the DFT data are shown in Fig.~\ref{fig:NN_Energy}(b)-(f). Compared to the linear EPI results, it is easy to note that the neural networks produce very high training scores, which demonstrates the superior representation capability of DNN. The testing scores are reasonably good, but generally worse than the other methods. This can also be seen from Fig.~\ref{fig:Model_Comparison}, where a comparison of the RMSEs from the four different models are shown. The relatively large difference between the training and testing errors of the neural network model indicates that the results can be improved by adding more data sets, giving prediction errors closer to the other models. 


Finally, we would like to point out that, while the neural network model did not show better prediction accuracy for the materials studied, the results are still very encouraging because they demonstrated that with sufficiently large data sets, a direct mapping from the local chemical environment to the atomic local energy is feasible. The versatile DNN models are of great potential to describe materials where non-linear atomic interactions are important. These interactions are difficult to include in the traditional methods, but straightforward for neural networks since they work in the configuration space and uses nonlinear activation functions.

In conclusion, we developed an approach to obtain large amounts of DFT data sets by employing the $O(N)$ LSMS method to calculate the atomic local energies. Such a method reduces the time scaling of generating a single data point to $O(0)$, which not only substantially speeds up the construction of effective Hamiltonian, but also allows for the use of complex models to describe non-linear atomic interactions. Using the large DFT data sets, a range of refractory HEAs were studied with four different models, among them the quadratic EPI model demonstrated the best overall performance. Neural networks were successfully applied to establish a mapping directly from the local chemical environment to the atomic local energy, which provides a promising tool for the future study of complex alloys.

\section{Method}
The DFT data are calculated with the locally self-consistent multiple scattering (LSMS) method \cite{PhysRevLett.75.2867}, which is a real space implementation of the Korringa-Kohn-Rostoker (KKR) method \cite{KORRINGA1947392, PhysRev.94.1111} and scales linearly with respect to the number atoms. The angular momentum cutoff in the LSMS method is set as 3. The local interaction zone (LIZ) with 59 atoms is found to be accurate enough and adopted after testing with both KKR method and LSMS method with LIZ = 283. Note that the LIZ in the LSMS only restricts the electron quantum scattering, while the LIR in the effective Hamiltonian is a cutoff on all the atomic interactions. The scalar-relativistic equations are solved to properly treat the heavier elements in the system. For simplicity, the atoms are
assumed to be on perfect BCC lattice sites. For MoNbTaW this was found to only affect the low temperature ground states \cite{Korman_npj}. 

The MoNbTaW data sets for the linear EPI model are calculated with supercells of 64, 128, and 256 atoms. 10 chemical configurations are generated randomly for each supercell, therefore the total number of training data is 4,480, with 1,120 data for each element. For the 5-element HEAs, the data sets for the linear EPI model are calculated with supercells of 20, 40, 80, and 160 atoms. 10 different configurations are generated randomly for each supercell, so the total number of training data sets is 3,000, with 600 data for each element. The same number of testing data are calculated using configurations different from the training ones.

More data sets are used for the other three models. For MoNbTaW, supercells of 64, 128, 256, and 512 atoms are used, giving a total of 38,400 training data and the same number of testing data. For MoNbTaVW, supercells of 20, 40, 80, 160, 320, 640, and 1280 atoms are used, giving a total of 50,800 training data and the same number of testing data. For MoNbTaTiW, supercells of 20, 40, 80, 160, and 320 atoms are used, giving a total of 62,000 training data and 29,840 testing data. For AlMoNbTaW, supercells of 20, 40, 80, and 160 atoms are used, giving a total of 30,000 training data and the same number of testing data.

The Scikit-learn \cite{scikit-learn} package is used for the linear and quadratic regression models. PyTorch is used for the neural network models, where mini-batch gradient descent, adaptive moment estimation (Adam) \cite{2014arXiv1412.6980K}, and back-propagation algorithms are employed to update the neural network parameters. The mini-batch size is chosen as 16 and the learning rate is set as 0.001. The loss function is simply the energy mean square error (MSE), with an additional L2 regularization penalty of $\alpha=1\times 10^{-8}$. 

\section{Data availability}
The data that support the findings of this study are available from the corresponding author upon request.

\section{Code availability}
The data that support the findings of this study are available from the corresponding author upon request.

\section{Acknowledgements}
The work of X. L. and M. E. was supported by the U.S. Department of Energy, Office of Science, Basic Energy Sciences, Materials Science and Engineering Division. J. Z. was supported by the Laboratory Directed Research and 
Development Program of Oak Ridge National Laboratory. This research used resources of the Oak Ridge Leadership Computing Facility, which is supported by the Office of Science of the U.S. Department of Energy under Contract No. DE-AC05-00OR22725. 

\section{Contributions}
X. L. conceived the project, calculated the data, and wrote the first draft. X. L. and J. Z. analyzed the results. Y. W. and M. E. developed the LSMS code. Y. W. modified the LSMS code to calculate the atomic local energy. All authors contributed to  the final manuscript.

\section{Competing interests}
The authors declare no competing interests.

\bibliographystyle{model1-num-names}
\bibliography{sample.bib}

\begin{thebibliography}{44}
\expandafter\ifx\csname natexlab\endcsname\relax\def\natexlab#1{#1}\fi
\providecommand{\bibinfo}[2]{#2}
\ifx\xfnm\relax \def\xfnm[#1]{\unskip,\space#1}\fi
\bibitem[{Widom(2018)}]{widom_2018}
\bibinfo{author}{M.~Widom},
\newblock \bibinfo{title}{Modeling the structure and thermodynamics of
  high-entropy alloys},
\newblock \bibinfo{journal}{Journal of Materials Research} \bibinfo{volume}{33}
  (\bibinfo{year}{2018}) \bibinfo{pages}{2881–2898}.
\bibitem[{Eisenbach et~al.(2019)Eisenbach, Pei, and Liu}]{Eisenbach_2019}
\bibinfo{author}{M.~Eisenbach}, \bibinfo{author}{Z.~Pei},
  \bibinfo{author}{X.~Liu},
\newblock \bibinfo{title}{First-principles study of order-disorder transitions
  in multicomponent solid-solution alloys},
\newblock \bibinfo{journal}{Journal of Physics: Condensed Matter}
  \bibinfo{volume}{31} (\bibinfo{year}{2019}) \bibinfo{pages}{273002}.
\bibitem[{Fern{\'a}ndez-Caballero et~al.(2017)Fern{\'a}ndez-Caballero,
  Wr{\'o}bel, Mummery, and Nguyen-Manh}]{Fernandez-Caballero2017}
\bibinfo{author}{A.~Fern{\'a}ndez-Caballero}, \bibinfo{author}{J.~S.
  Wr{\'o}bel}, \bibinfo{author}{P.~M. Mummery},
  \bibinfo{author}{D.~Nguyen-Manh},
\newblock \bibinfo{title}{Short-range order in high entropy alloys: Theoretical
  formulation and application to {Mo-Nb-Ta-V-W} system},
\newblock \bibinfo{journal}{Journal of Phase Equilibria and Diffusion}
  \bibinfo{volume}{38} (\bibinfo{year}{2017}) \bibinfo{pages}{391--403}.
\bibitem[{Khan and Eisenbach(2016)}]{PhysRevB.93.024203}
\bibinfo{author}{S.~N. Khan}, \bibinfo{author}{M.~Eisenbach},
\newblock \bibinfo{title}{{Density-functional Monte-Carlo simulation of CuZn
  order-disorder transition}},
\newblock \bibinfo{journal}{Phys. Rev. B} \bibinfo{volume}{93}
  (\bibinfo{year}{2016}) \bibinfo{pages}{024203}.
\bibitem[{Kikuchi(1951)}]{PhysRev.81.988}
\bibinfo{author}{R.~Kikuchi},
\newblock \bibinfo{title}{A theory of cooperative phenomena},
\newblock \bibinfo{journal}{Phys. Rev.} \bibinfo{volume}{81}
  (\bibinfo{year}{1951}) \bibinfo{pages}{988--1003}.
\bibitem[{Sanchez et~al.(1984)Sanchez, Ducastelle, and
  Gratias}]{SANCHEZ1984334}
\bibinfo{author}{J.~Sanchez}, \bibinfo{author}{F.~Ducastelle},
  \bibinfo{author}{D.~Gratias},
\newblock \bibinfo{title}{Generalized cluster description of multicomponent
  systems},
\newblock \bibinfo{journal}{Physica A: Statistical Mechanics and its
  Applications} \bibinfo{volume}{128} (\bibinfo{year}{1984})
  \bibinfo{pages}{334 -- 350}.
\bibitem[{Feng et~al.(2017)Feng, Liaw, Gao, and Widom}]{NPJ_Widom}
\bibinfo{author}{R.~Feng}, \bibinfo{author}{P.~K. Liaw}, \bibinfo{author}{M.~C.
  Gao}, \bibinfo{author}{M.~Widom},
\newblock \bibinfo{title}{First-principles prediction of high-entropy-alloy
  stability},
\newblock \bibinfo{journal}{npj Computational Materials} \bibinfo{volume}{3}
  (\bibinfo{year}{2017}) \bibinfo{pages}{50}.
\bibitem[{Blum et~al.(2005)Blum, Hart, Walorski, and
  Zunger}]{PhysRevB.72.165113}
\bibinfo{author}{V.~Blum}, \bibinfo{author}{G.~L.~W. Hart},
  \bibinfo{author}{M.~J. Walorski}, \bibinfo{author}{A.~Zunger},
\newblock \bibinfo{title}{Using genetic algorithms to map first-principles
  results to model hamiltonians: Application to the generalized ising model for
  alloys},
\newblock \bibinfo{journal}{Phys. Rev. B} \bibinfo{volume}{72}
  (\bibinfo{year}{2005}) \bibinfo{pages}{165113}.
\bibitem[{Levy et~al.(2010)Levy, Hart, and Curtarolo}]{doi:10.1021/ja9105623}
\bibinfo{author}{O.~Levy}, \bibinfo{author}{G.~L.~W. Hart},
  \bibinfo{author}{S.~Curtarolo},
\newblock \bibinfo{title}{Uncovering compounds by synergy of cluster expansion
  and high-throughput methods},
\newblock \bibinfo{journal}{Journal of the American Chemical Society}
  \bibinfo{volume}{132} (\bibinfo{year}{2010}) \bibinfo{pages}{4830--4833}.
  \bibinfo{note}{PMID: 20218599}.
\bibitem[{Ruban and Abrikosov(2008)}]{RubanECIReview}
\bibinfo{author}{A.~V. Ruban}, \bibinfo{author}{I.~A. Abrikosov},
\newblock \bibinfo{title}{Configurational thermodynamics of alloys from first
  principles: effective cluster interactions},
\newblock \bibinfo{journal}{Rep. Prog. Phys.} \bibinfo{volume}{71}
  (\bibinfo{year}{2008}) \bibinfo{pages}{046501}.
\bibitem[{Connolly and Williams(1983)}]{PhysRevB.27.5169}
\bibinfo{author}{J.~W.~D. Connolly}, \bibinfo{author}{A.~R. Williams},
\newblock \bibinfo{title}{Density-functional theory applied to phase
  transformations in transition-metal alloys},
\newblock \bibinfo{journal}{Phys. Rev. B} \bibinfo{volume}{27}
  (\bibinfo{year}{1983}) \bibinfo{pages}{5169--5172}.
\bibitem[{Jiang and Uberuaga(2016)}]{PhysRevLett.116.105501}
\bibinfo{author}{C.~Jiang}, \bibinfo{author}{B.~P. Uberuaga},
\newblock \bibinfo{title}{Efficient ab initio modeling of random multicomponent
  alloys},
\newblock \bibinfo{journal}{Phys. Rev. Lett.} \bibinfo{volume}{116}
  (\bibinfo{year}{2016}) \bibinfo{pages}{105501}.
\bibitem[{Seko et~al.(2009)Seko, Koyama, and Tanaka}]{seko2009cluster}
\bibinfo{author}{A.~Seko}, \bibinfo{author}{Y.~Koyama},
  \bibinfo{author}{I.~Tanaka},
\newblock \bibinfo{title}{Cluster expansion method for multicomponent systems
  based on optimal selection of structures for density-functional theory
  calculations},
\newblock \bibinfo{journal}{Physical Review B} \bibinfo{volume}{80}
  (\bibinfo{year}{2009}) \bibinfo{pages}{165122}.
\bibitem[{Yeh et~al.(2004)Yeh, Chen, Lin, Gan, Chin, Shun, Tsau, and
  Chang}]{ADEM:ADEM200300567}
\bibinfo{author}{J.-W. Yeh}, \bibinfo{author}{S.-K. Chen},
  \bibinfo{author}{S.-J. Lin}, \bibinfo{author}{J.-Y. Gan},
  \bibinfo{author}{T.-S. Chin}, \bibinfo{author}{T.-T. Shun},
  \bibinfo{author}{C.-H. Tsau}, \bibinfo{author}{S.-Y. Chang},
\newblock \bibinfo{title}{Nanostructured high-entropy alloys with multiple
  principal elements: Novel alloy design concepts and outcomes},
\newblock \bibinfo{journal}{Advanced Engineering Materials} \bibinfo{volume}{6}
  (\bibinfo{year}{2004}) \bibinfo{pages}{299--303}.
\bibitem[{Cantor et~al.(2004)Cantor, Chang, Knight, and
  Vincent}]{CANTOR2004213}
\bibinfo{author}{B.~Cantor}, \bibinfo{author}{I.~Chang},
  \bibinfo{author}{P.~Knight}, \bibinfo{author}{A.~Vincent},
\newblock \bibinfo{title}{Microstructural development in equiatomic
  multicomponent alloys},
\newblock \bibinfo{journal}{Materials Science and Engineering: A}
  \bibinfo{volume}{375-377} (\bibinfo{year}{2004}) \bibinfo{pages}{213 -- 218}.
\bibitem[{George et~al.(2019)George, Raabe, and Ritchie}]{George2019}
\bibinfo{author}{E.~P. George}, \bibinfo{author}{D.~Raabe},
  \bibinfo{author}{R.~O. Ritchie},
\newblock \bibinfo{title}{High-entropy alloys},
\newblock \bibinfo{journal}{Nature Reviews Materials} \bibinfo{volume}{4}
  (\bibinfo{year}{2019}) \bibinfo{pages}{515--534}.
\bibitem[{Gludovatz et~al.(2016)Gludovatz, Hohenwarter, Thurston, Bei, Wu,
  George, and Ritchie}]{NatureComNiCoCr}
\bibinfo{author}{B.~Gludovatz}, \bibinfo{author}{A.~Hohenwarter},
  \bibinfo{author}{K.~V.~S. Thurston}, \bibinfo{author}{H.~Bei},
  \bibinfo{author}{Z.~Wu}, \bibinfo{author}{E.~P. George},
  \bibinfo{author}{R.~O. Ritchie},
\newblock \bibinfo{title}{{Exceptional damage-tolerance of a medium-entropy
  alloy {CrCoNi} at cryogenic temperatures}},
\newblock \bibinfo{journal}{Nature Communications} \bibinfo{volume}{7}
  (\bibinfo{year}{2016}) \bibinfo{pages}{10602}.
\bibitem[{Gludovatz et~al.(2014)Gludovatz, Hohenwarter, Catoor, Chang, George,
  and Ritchie}]{Gludovatz1153}
\bibinfo{author}{B.~Gludovatz}, \bibinfo{author}{A.~Hohenwarter},
  \bibinfo{author}{D.~Catoor}, \bibinfo{author}{E.~H. Chang},
  \bibinfo{author}{E.~P. George}, \bibinfo{author}{R.~O. Ritchie},
\newblock \bibinfo{title}{A fracture-resistant high-entropy alloy for cryogenic
  applications},
\newblock \bibinfo{journal}{Science} \bibinfo{volume}{345}
  (\bibinfo{year}{2014}) \bibinfo{pages}{1153--1158}.
\bibitem[{Senkov et~al.(2011)Senkov, Wilks, Scott, and Miracle}]{SENKOV2011698}
\bibinfo{author}{O.~Senkov}, \bibinfo{author}{G.~Wilks},
  \bibinfo{author}{J.~Scott}, \bibinfo{author}{D.~Miracle},
\newblock \bibinfo{title}{{Mechanical properties of Nb25Mo25Ta25W25 and
  V20Nb20Mo20Ta20W20 refractory high entropy alloys}},
\newblock \bibinfo{journal}{Intermetallics} \bibinfo{volume}{19}
  (\bibinfo{year}{2011}) \bibinfo{pages}{698 -- 706}.
\bibitem[{Miracle and Senkov(2017)}]{MIRACLE2017448}
\bibinfo{author}{D.~Miracle}, \bibinfo{author}{O.~Senkov},
\newblock \bibinfo{title}{A critical review of high entropy alloys and related
  concepts},
\newblock \bibinfo{journal}{Acta Materialia} \bibinfo{volume}{122}
  (\bibinfo{year}{2017}) \bibinfo{pages}{448 -- 511}.
\bibitem[{Gyorffy and Stocks(1983)}]{PhysRevLett.50.374}
\bibinfo{author}{B.~L. Gyorffy}, \bibinfo{author}{G.~M. Stocks},
\newblock \bibinfo{title}{Concentration waves and fermi surfaces in random
  metallic alloys},
\newblock \bibinfo{journal}{Phys. Rev. Lett.} \bibinfo{volume}{50}
  (\bibinfo{year}{1983}) \bibinfo{pages}{374--377}.
\bibitem[{Ducastelle and Gautier(1976)}]{0305-4608-6-11-005}
\bibinfo{author}{F.~Ducastelle}, \bibinfo{author}{F.~Gautier},
\newblock \bibinfo{title}{Generalized perturbation theory in disordered
  transitional alloys: Applications to the calculation of ordering energies},
\newblock \bibinfo{journal}{Journal of Physics F: Metal Physics}
  \bibinfo{volume}{6} (\bibinfo{year}{1976}) \bibinfo{pages}{2039}.
\bibitem[{Gonis et~al.(1987)Gonis, Zhang, Freeman, Turchi, Stocks, and
  Nicholson}]{PhysRevB.36.4630}
\bibinfo{author}{A.~Gonis}, \bibinfo{author}{X.~G. Zhang},
  \bibinfo{author}{A.~J. Freeman}, \bibinfo{author}{P.~Turchi},
  \bibinfo{author}{G.~M. Stocks}, \bibinfo{author}{D.~M. Nicholson},
\newblock \bibinfo{title}{Configurational energies and effective cluster
  interactions in substitutionally disordered binary alloys},
\newblock \bibinfo{journal}{Phys. Rev. B} \bibinfo{volume}{36}
  (\bibinfo{year}{1987}) \bibinfo{pages}{4630--4646}.
\bibitem[{Natarajan and Van~der Ven(2018)}]{ML_Cluster}
\bibinfo{author}{A.~R. Natarajan}, \bibinfo{author}{A.~Van~der Ven},
\newblock \bibinfo{title}{Machine-learning the configurational energy of
  multicomponent crystalline solids},
\newblock \bibinfo{journal}{npj Computational Materials} \bibinfo{volume}{4}
  (\bibinfo{year}{2018}) \bibinfo{pages}{56}.
\bibitem[{{Gao} et~al.(2015){Gao}, {Yao}, {Schneider}, and
  {Widom}}]{2015arXiv151209110G}
\bibinfo{author}{Q.~{Gao}}, \bibinfo{author}{S.~{Yao}},
  \bibinfo{author}{J.~{Schneider}}, \bibinfo{author}{M.~{Widom}},
\newblock \bibinfo{title}{{Machine Learning methods for interatomic potentials:
  application to boron carbide}},
\newblock \bibinfo{journal}{arXiv e-prints}  (\bibinfo{year}{2015})
  \bibinfo{pages}{arXiv:1512.09110}.
\bibitem[{LeCun et~al.(2015)LeCun, Bengio, and Hinton}]{NatureDL}
\bibinfo{author}{Y.~LeCun}, \bibinfo{author}{Y.~Bengio},
  \bibinfo{author}{G.~Hinton},
\newblock \bibinfo{title}{Deep learning},
\newblock \bibinfo{journal}{Nature} \bibinfo{volume}{521}
  (\bibinfo{year}{2015}) \bibinfo{pages}{436}.
\bibitem[{Chmiela et~al.(2017)Chmiela, Tkatchenko, Sauceda, Poltavsky,
  Sch{\"u}tt, and M{\"u}ller}]{Chmielae1603015}
\bibinfo{author}{S.~Chmiela}, \bibinfo{author}{A.~Tkatchenko},
  \bibinfo{author}{H.~E. Sauceda}, \bibinfo{author}{I.~Poltavsky},
  \bibinfo{author}{K.~T. Sch{\"u}tt}, \bibinfo{author}{K.-R. M{\"u}ller},
\newblock \bibinfo{title}{Machine learning of accurate energy-conserving
  molecular force fields},
\newblock \bibinfo{journal}{Science Advances} \bibinfo{volume}{3}
  (\bibinfo{year}{2017}).
\bibitem[{Li et~al.(2015)Li, Kermode, and De~Vita}]{PhysRevLett.114.096405}
\bibinfo{author}{Z.~Li}, \bibinfo{author}{J.~R. Kermode},
  \bibinfo{author}{A.~De~Vita},
\newblock \bibinfo{title}{Molecular dynamics with on-the-fly machine learning
  of quantum-mechanical forces},
\newblock \bibinfo{journal}{Phys. Rev. Lett.} \bibinfo{volume}{114}
  (\bibinfo{year}{2015}) \bibinfo{pages}{096405}.
\bibitem[{Deringer and Cs\'anyi(2017)}]{PhysRevB.95.094203}
\bibinfo{author}{V.~L. Deringer}, \bibinfo{author}{G.~Cs\'anyi},
\newblock \bibinfo{title}{Machine learning based interatomic potential for
  amorphous carbon},
\newblock \bibinfo{journal}{Phys. Rev. B} \bibinfo{volume}{95}
  (\bibinfo{year}{2017}) \bibinfo{pages}{094203}.
\bibitem[{Bart\'ok et~al.(2018)Bart\'ok, Kermode, Bernstein, and
  Cs\'anyi}]{PhysRevX.8.041048}
\bibinfo{author}{A.~P. Bart\'ok}, \bibinfo{author}{J.~Kermode},
  \bibinfo{author}{N.~Bernstein}, \bibinfo{author}{G.~Cs\'anyi},
\newblock \bibinfo{title}{Machine learning a general-purpose interatomic
  potential for silicon},
\newblock \bibinfo{journal}{Phys. Rev. X} \bibinfo{volume}{8}
  (\bibinfo{year}{2018}) \bibinfo{pages}{041048}.
\bibitem[{Kostiuchenko et~al.(2019)Kostiuchenko, K\"ormann, Neugebauer, and
  Shapeev}]{Korman_npj}
\bibinfo{author}{T.~Kostiuchenko}, \bibinfo{author}{F.~K\"ormann},
  \bibinfo{author}{J.~Neugebauer}, \bibinfo{author}{A.~Shapeev},
\newblock \bibinfo{title}{Impact of lattice relaxations on phase transitions in
  a high-entropy alloy studied by machine-learning potentials},
\newblock \bibinfo{journal}{npj Computational Materials} \bibinfo{volume}{5}
  (\bibinfo{year}{2019}) \bibinfo{pages}{55}.
\bibitem[{Ye et~al.(2018)Ye, Chen, Wang, Chu, and Ong}]{Ye2018}
\bibinfo{author}{W.~Ye}, \bibinfo{author}{C.~Chen}, \bibinfo{author}{Z.~Wang},
  \bibinfo{author}{I.-H. Chu}, \bibinfo{author}{S.~P. Ong},
\newblock \bibinfo{title}{Deep neural networks for accurate predictions of
  crystal stability},
\newblock \bibinfo{journal}{Nature Communications} \bibinfo{volume}{9}
  (\bibinfo{year}{2018}) \bibinfo{pages}{3800}.
\bibitem[{MEH(2019)}]{MEHTA20191}
\bibinfo{title}{A high-bias, low-variance introduction to machine learning for
  physicists},
\newblock \bibinfo{journal}{Physics Reports} \bibinfo{volume}{810}
  (\bibinfo{year}{2019}) \bibinfo{pages}{1 -- 124}.
\bibitem[{Goodfellow et~al.(2016)Goodfellow, Bengio, and
  Courville}]{Goodfellow-et-al-2016}
\bibinfo{author}{I.~Goodfellow}, \bibinfo{author}{Y.~Bengio},
  \bibinfo{author}{A.~Courville}, \bibinfo{title}{Deep Learning},
  \bibinfo{publisher}{MIT Press}, \bibinfo{year}{2016}.
\bibitem[{Goedecker(1999)}]{RevModPhys.71.1085}
\bibinfo{author}{S.~Goedecker},
\newblock \bibinfo{title}{Linear scaling electronic structure methods},
\newblock \bibinfo{journal}{Rev. Mod. Phys.} \bibinfo{volume}{71}
  (\bibinfo{year}{1999}) \bibinfo{pages}{1085--1123}.
\bibitem[{Prodan and Kohn(2005)}]{Prodan11635}
\bibinfo{author}{E.~Prodan}, \bibinfo{author}{W.~Kohn},
\newblock \bibinfo{title}{Nearsightedness of electronic matter},
\newblock \bibinfo{journal}{Proceedings of the National Academy of Sciences}
  \bibinfo{volume}{102} (\bibinfo{year}{2005}) \bibinfo{pages}{11635--11638}.
\bibitem[{Wang et~al.(1995)Wang, Stocks, Shelton, Nicholson, Szotek, and
  Temmerman}]{PhysRevLett.75.2867}
\bibinfo{author}{Y.~Wang}, \bibinfo{author}{G.~M. Stocks},
  \bibinfo{author}{W.~A. Shelton}, \bibinfo{author}{D.~M.~C. Nicholson},
  \bibinfo{author}{Z.~Szotek}, \bibinfo{author}{W.~M. Temmerman},
\newblock \bibinfo{title}{Order-{N} multiple scattering approach to electronic
  structure calculations},
\newblock \bibinfo{journal}{Phys. Rev. Lett.} \bibinfo{volume}{75}
  (\bibinfo{year}{1995}) \bibinfo{pages}{2867--2870}.
\bibitem[{Huhn and Widom(2013)}]{Huhn2013}
\bibinfo{author}{W.~P. Huhn}, \bibinfo{author}{M.~Widom},
\newblock \bibinfo{title}{Prediction of {A2 to B2} phase transition in the
  high-entropy alloy {Mo-Nb-Ta-W}},
\newblock \bibinfo{journal}{JOM} \bibinfo{volume}{65} (\bibinfo{year}{2013})
  \bibinfo{pages}{1772--1779}.
\bibitem[{Zhang et~al.(2020)Zhang, Liu, Bi, Yin, Zhang, and
  Eisenbach}]{ZHANG2020108247}
\bibinfo{author}{J.~Zhang}, \bibinfo{author}{X.~Liu}, \bibinfo{author}{S.~Bi},
  \bibinfo{author}{J.~Yin}, \bibinfo{author}{G.~Zhang},
  \bibinfo{author}{M.~Eisenbach},
\newblock \bibinfo{title}{Robust data-driven approach for predicting the
  configurational energy of high entropy alloys},
\newblock \bibinfo{journal}{Materials \& Design} \bibinfo{volume}{185}
  (\bibinfo{year}{2020}) \bibinfo{pages}{108247}.
\bibitem[{Leshno et~al.(1993)Leshno, Lin, Pinkus, and Schocken}]{LESHNO1993861}
\bibinfo{author}{M.~Leshno}, \bibinfo{author}{V.~Y. Lin},
  \bibinfo{author}{A.~Pinkus}, \bibinfo{author}{S.~Schocken},
\newblock \bibinfo{title}{Multilayer feedforward networks with a nonpolynomial
  activation function can approximate any function},
\newblock \bibinfo{journal}{Neural Networks} \bibinfo{volume}{6}
  (\bibinfo{year}{1993}) \bibinfo{pages}{861 -- 867}.
\bibitem[{Korringa(1947)}]{KORRINGA1947392}
\bibinfo{author}{J.~Korringa},
\newblock \bibinfo{title}{On the calculation of the energy of a {Bloch} wave in
  a metal},
\newblock \bibinfo{journal}{Physica} \bibinfo{volume}{13}
  (\bibinfo{year}{1947}) \bibinfo{pages}{392 -- 400}.
\bibitem[{Kohn and Rostoker(1954)}]{PhysRev.94.1111}
\bibinfo{author}{W.~Kohn}, \bibinfo{author}{N.~Rostoker},
\newblock \bibinfo{title}{Solution of the {Schr\"odinger} equation in periodic
  lattices with an application to metallic lithium},
\newblock \bibinfo{journal}{Phys. Rev.} \bibinfo{volume}{94}
  (\bibinfo{year}{1954}) \bibinfo{pages}{1111--1120}.
\bibitem[{Pedregosa et~al.(2011)Pedregosa, Varoquaux, Gramfort, Michel,
  Thirion, Grisel, Blondel, Prettenhofer, Weiss, Dubourg, Vanderplas, Passos,
  Cournapeau, Brucher, Perrot, and Duchesnay}]{scikit-learn}
\bibinfo{author}{F.~Pedregosa}, \bibinfo{author}{G.~Varoquaux},
  \bibinfo{author}{A.~Gramfort}, \bibinfo{author}{V.~Michel},
  \bibinfo{author}{B.~Thirion}, \bibinfo{author}{O.~Grisel},
  \bibinfo{author}{M.~Blondel}, \bibinfo{author}{P.~Prettenhofer},
  \bibinfo{author}{R.~Weiss}, \bibinfo{author}{V.~Dubourg},
  \bibinfo{author}{J.~Vanderplas}, \bibinfo{author}{A.~Passos},
  \bibinfo{author}{D.~Cournapeau}, \bibinfo{author}{M.~Brucher},
  \bibinfo{author}{M.~Perrot}, \bibinfo{author}{E.~Duchesnay},
\newblock \bibinfo{title}{Scikit-learn: Machine learning in {P}ython},
\newblock \bibinfo{journal}{Journal of Machine Learning Research}
  \bibinfo{volume}{12} (\bibinfo{year}{2011}) \bibinfo{pages}{2825--2830}.
\bibitem[{{Kingma} and {Ba}(2014)}]{2014arXiv1412.6980K}
\bibinfo{author}{D.~P. {Kingma}}, \bibinfo{author}{J.~{Ba}},
\newblock \bibinfo{title}{{Adam: A Method for Stochastic Optimization}},
\newblock \bibinfo{journal}{arXiv e-prints}  (\bibinfo{year}{2014})
  \bibinfo{pages}{arXiv:1412.6980}.

\end{thebibliography}

\end{document}